\documentclass{aa}  

\usepackage{graphicx}
\usepackage{txfonts}
\usepackage{lipsum}
\usepackage{subcaption}         
\usepackage{lscape}             
\usepackage{placeins}           
\usepackage{float}
\usepackage{multirow}
\usepackage{amsmath,amssymb,amsfonts}
\usepackage{mathrsfs}
\usepackage{xcolor}
\usepackage{booktabs}
\usepackage{url}
\usepackage{enumerate}
\usepackage{siunitx}
\usepackage{array}    
\usepackage{dcolumn}  
\usepackage{twemojis}
\usepackage{float}
\usepackage[colorlinks=true, allcolors=blue]{hyperref}

\RequirePackage{etex}

\newcommand{\gaia}{\emph{Gaia}}

\newcommand{\gex}{\texttt{GEX}}
\newcommand{\tgex}{\texttt{T-GEX}}
\newcommand{\cachai}{\textsc{Cachai}}

\newcommand{\feh}{\ensuremath{[\mathrm{Fe}/\mathrm{H}]}}
\newcommand{\mgfe}{\ensuremath{[\mathrm{Mg}/\mathrm{Fe}]}}
\newcommand{\mgh}{\ensuremath{[\mathrm{Mg}/\mathrm{H}]}}
\newcommand{\teff}{\ensuremath{T_{\rm eff}}}
\newcommand{\logg}{\ensuremath{\log g}}

\newcommand{\zmax}{$\langle Z_{\rm max} \rangle$}
\newcommand{\lz}{$\langle L_z \rangle$}
\newcommand{\afe}{[$\alpha$/Fe]}
\newcommand{\rb}{$R_{\rm b}$}
\newcommand{\rgui}{$\langle R_{\rm g} \rangle$}
\newcommand{\fuvnuv}{FUV--NUV}
\newcommand{\nuvg}{NUV--$G$}

\defcitealias{Dantas2026_TGEXI}{Paper I}
\titlerunning{\reflectbox{\twemoji[scale=0.5]{t-rex}} The \tgex\ project}

\begin{document}

   \title{\reflectbox{\twemoji[scale=0.8]{t-rex}} The \tgex\ project}

   \subtitle{II. Chrono-chemo-dynamic signatures of UV-bright FGK stars}

%
%
%

    \author{
              D. Beltrán\inst{\ref{aff:puc_ia}, \dagger,}\corrauth{d\_beltran@uc.cl}
              \and
              M. L. L. Dantas\inst{\ref{aff:puc_ia}, \dagger}\corrauth{mlldantas@protonmail.com; mlldantas@uc.cl}
              \and
              R. Smiljanic\inst{\ref{aff:camk}}
              \and
              P. B. Tissera\inst{\ref{aff:puc_ia}}
              }

   \institute{
              Instituto de Astrofísica, Pontificia Universidad Católica de Chile, Av. Vicuña Mackenna 4860, Santiago, Chile \label{aff:puc_ia}
              \and
              Nicolaus Copernicus Astronomical Center, Polish Academy of Sciences, ul. Bartycka 18, 00-716, Warsaw, Poland \label{aff:camk}
             }

   \date{Received Month Day, 20XX}

   \abstract
   {The origin of anomalous ultraviolet (UV) emission in apparently unevolved FGK stars remains unclear. Chemo-dynamical differences among stars with distinct UV excess levels may constrain its source and connection to the UV upturn in unresolved populations.}
   {To probe whether these groups of stars differ in intrinsic or kinematic stellar properties, we compare the ages, chemical abundances, Galactic orbits, and inferred birth environments (for thin disc members) of 37 \tgex\ stars with strong (G1), normal (G2), and intermediate (G3) UV emission.}
   {We constructed group-wise chord diagrams to investigate correlations; integrated Galactic orbits from \gaia\ DR3 astrometry; examined chemo-dynamical diagrams; inferred birth radii for probable thin disc members; and hierarchically clustered individual abundance profiles.}
   {The groups show distinct correlation structures: UV colours correlate with chemical and orbital quantities in G1; G2 has the largest number of significant (anti-)correlations; and no UV-colour correlation reaches the adopted threshold in G3. Most stars have prograde disc-like orbits, while both halo-like objects belong to G2. The groups overlap in the \mgfe--\feh\ plane, but G1 is shifted towards lower \feh\ and higher \mgfe. Among the 18 probable thin-disc members, G3 has a smaller median birth radius than G2 ($5.70$ versus $8.10$~kpc), despite their nearly identical median guiding radii (\rgui; $8.23$ versus $8.25$~kpc). All five G1 stars are Ti-enriched compared to Fe, and four combine enhanced Co with comparatively low Mn and, in some cases, Cr. A related but less complete pattern occurs in G3, whereas Co enhancement in some G2 stars does not reproduce the joint signature.}
   {The UV-defined groups are heterogeneous and do not correspond to unique Galactic components. The convergence of $\alpha$-enhancement (likely thick disc members), positive Ti ratios, and the Cr--Mn--Fe--Co--Ni chemical abundance pattern (clearest in G1 but present in related form in G3) supports high-energy core-collapse (i.e. hypernovae) enrichment of the natal material of some UV-excess stars. Additionally, the smaller G3 birth radii (for those in the Galactic thin disc) indicate more inward formation regions and provide a possible spatial link to UV-bright stellar populations in the bulges of spiral galaxies. Neither result establishes a single causal pathway. Whether mixture-dependent metal-line blanketing connects this enrichment to the UV emission requires tailored spectral modelling.}


   \maketitle
   \nolinenumbers

   \begingroup
   \renewcommand{\thefootnote}{\ensuremath{\dagger}}
   \footnotetext{Contributed equally to this work.}
   \endgroup

\section{Introduction}
\label{sec:intro}

The Milky Way (MW) offers a uniquely detailed view of the processes that shape galaxies across cosmic time \citep[e.g.][]{RixBovy2013, Helmi2020}. While the stellar populations of external galaxies are largely observed through their integrated light \citep[][]{Walcher2011, Conroy2013, Coelho2020}, the Galaxy can be dissected star by star, allowing stellar ages, chemical abundances, kinematics, and orbital properties to be examined together. It therefore provides a fundamental laboratory in which evolutionary histories otherwise blurred together in unresolved systems can be reconstructed from the present-day properties of individual stars \citep{FBH2002, BHG2016, Helmi2020}.

This possibility lies at the heart of Galactic archaeology \citep[e.g.][]{FBH2002, BHG2016}. For long-lived stars, many elemental abundances preserve information about the interstellar medium from which they formed, while their ages and present-day phase-space coordinates constrain when they formed and how their orbits subsequently evolved. Together with models of Galactic chemical evolution, these complementary dimensions can also provide information about their likely birth environments. Their combination has revealed that the MW disc is neither chemically nor dynamically static \citep[e.g.][]{SellwoodBinney2002, RixBovy2013, Minchev2013, MartinezMedina2016, Frankel2018, Magrini2023, Dantas2023, Dantas2025a, Dantas2025b, GarciaDelgado2026}. Instead, its present-day stellar populations reflect the coupled effects of chemical enrichment, star formation, mergers, secular evolution, and radial redistribution over several Gyr \citep[][]{Chiappini1997, Chiappini2001, Carr2022}. Consequently, stars currently found in the same Galactic region need not share the same birth environment or evolutionary history \citep[e.g.][and references therein]{Trevisan2011, Minchev2018, Chen2019, ChenZhao2020, Feltzing2020, MartinezBautista2021, Okalidis2022, Ratcliffe2023, Dantas2023, Lu2024, Ratcliffe2025}.

The Tiny \gaia-ESO + GALEX (\tgex) project represents the resolved-stellar counterpart of a broader effort to understand the origin and evolution of the ultraviolet (UV) excess in old stellar populations. Traditionally investigated through the integrated UV emission of old stellar systems, such as elliptical galaxies, the bulges of spiral galaxies, globular clusters, and even old open clusters \citep[e.g.][]{Burstein1988, Dorman1995, Oconnell1999, Buzzoni2012, Dalessandro2012, Schiavon2012, Goudfrooij2018, Goudfrooij2026}, the UV upturn encodes the contribution of stellar components that remain difficult to identify unambiguously in unresolved populations.

From this perspective, \citet{Dantas2020} investigated the incidence of the UV upturn as a function of redshift and stellar mass using the GAMA survey \citep[][]{Driver2009, Baldry2018}, coupled with observations from GALEX \citep[][]{Martin2005} and SDSS DR7 \citep{York2000, Abazajian2009}. They found that the fraction of systems with UV upturn increases with stellar mass and rises towards a maximum around redshift $\sim0.25$, followed by a decline that remains to be confirmed. Subsequently, using a de-biased sample matched in redshift and stellar mass, showed that UV-weak and UV-upturn systems share broadly similar properties and may not constitute fundamentally distinct galaxy populations. Nevertheless, the UV-upturn systems tend to be more metal-rich and to have evolved more passively, with narrower ranges of stellar-population properties and less extended star-formation histories \citep[see][]{Dantas2021}. These results point towards a phenomenon connected to the long-term evolution of old, chemically enriched populations, while also illustrating the limitations of determining its stellar origin from integrated light alone.

The leading explanations for the UV upturn in genuinely old stellar populations invoke a minority population of hot, low-mass, evolved stars. Extreme horizontal-branch (EHB) stars and their post-HB descendants, particularly AGB-manqué and post-early-AGB stars, remain hot for sufficiently long periods to dominate the far-UV emission, whereas the shorter-lived post-AGB phase is generally expected to provide a more limited contribution \citep[e.g.][]{Dorman1993, Dorman1995, Oconnell1999, Brown2000, Brown2003, Rosenfield2012}. How old populations produce enough EHB stars to account for the observed emission, however, remains uncertain. Proposed routes include enhanced mass loss along the red-giant branch, which leaves unusually thin H-rich envelopes; genuinely He-enhanced subpopulations, whose evolution favours hotter HB morphologies; and binary interactions. The latter can produce hot subdwarfs through stable mass transfer, common-envelope ejection, or the merger of two He white dwarfs \citep[e.g.][]{Han2007, Chung2011, Heber2016}. Binary population-synthesis calculations have shown that these evolutionary products can reproduce the UV properties of old stellar populations and the UV--optical colours of observed UV-upturn galaxies (\citealt{HPB2013}, \citeyear{HPB2014}), while more general binary-inclusive frameworks demonstrate that stripping and accretion products can alter the integrated UV spectra of evolved populations \citep{Eldridge2017, Stanway2018}. The distinction between these channels is important: He-enhanced models assume that the stars formed with higher initial He abundances, whereas envelope-stripped hot subdwarfs need not originate from He-enhanced populations.

Other sources can further complicate the interpretation of integrated UV emission. Because young stellar populations emit far more UV light per unit mass than old populations, even a small young component can dominate the UV spectrum while leaving only a weak imprint at optical wavelengths. UV-extended population models and combined NUV--optical spectroscopy indicate young mass fractions of only ${\sim}0.1$--$0.5$ per cent in massive early-type galaxies \citep{Vazdekis2016, SalvadorRusinol2020}. Moreover, \citet{Werle2020} detected contributions from populations younger than 1~Gyr in 17.5 per cent of galaxies satisfying commonly adopted photometric UV-upturn criteria, demonstrating that residual star formation can remain present even in nominally selected UV-upturn samples. Within the same modelling framework, the old, hot central stars of planetary nebulae and white dwarfs could reproduce the far-UV emission of many of the remaining systems \citep[e.g.][]{Parsons2016}, although this result depends strongly on the adopted post-AGB evolutionary prescription and additional EHB or binary products were still required in some cases \citep[e.g.][]{Brown2008, Akhil2024}. Hot white dwarfs acting as direct UV emitters should therefore be distinguished from mergers of He white dwarfs, which instead constitute a formation route for EHB stars \citep[e.g.][]{Han2020, HPB2014}. Residual star formation is likewise best regarded as a competing source of UV emission \citep[in the form of young hot stars, e.g.][]{Vazdekis2016, SalvadorRusinol2020}, rather than as a mechanism producing the genuine UV upturn of an exclusively old population. The degeneracy among these contributions reinforces the difficulty of identifying the responsible stars from unresolved observations alone.

A natural approach is therefore to search for and characterise possible contributors at the level of individual stars. Photometric studies have already revealed optically ordinary Galactic stars with unexpectedly strong or unusual UV emission in NUV, FUV, or both \citep[e.g.][]{Smith2014, Melis2025}, with proposed explanations including enhanced chromospheric activity and unresolved hot white-dwarf companions \citep[e.g.][]{Parsons2016, Nayak2024}. Because this approach requires the UV emission to be associated as reliably as possible with a single optical source, \tgex\ focuses on Galactic-field stars and avoids crowded environments such as globular clusters. At the several-arcsec resolution of GALEX, multiple optical sources may otherwise fall within the same UV resolution element, making source confusion and blending particularly difficult to exclude \citep{Morrissey2007, Bianchi2014}.

The combination of \gaia\ astrometry \citep{Gaia2016, GaiaEDR3, Gaia2023_DR3} with large spectroscopic surveys has made it possible to explore the chrono-chemo-dynamical properties of increasingly specialised stellar populations \citep[e.g.][]{Cui2012, Gilmore2012, Gilmore2022, Randich2013, Randich2022, Buder2025}. Such contextual information can be particularly valuable for stars whose observed properties are difficult to reconcile with conventional expectations. More generally, ages, abundance patterns, and orbital histories can help distinguish an unusual evolutionary pathway from a particular formation environment or the extreme tail of an otherwise ordinary Galactic population.

Elemental-abundance ratios may be especially relevant in the context of the UV upturn. At the galaxy scale, \citet{Carter2011} found that the strength of the FUV excess correlates with both \afe\ and $[Z/{\rm H}]$, and more strongly with the former. Although the same mechanism need not operate in individual UV-bright stars, this motivates examining their $\alpha$-element abundances, which can affect stellar atmospheres and populations synthesis spectra \citep[e.g.][]{Coelho2007, Coelho2014, Vazdekis2015}.

In the first paper of the \tgex\ project \citep[][hereafter \citetalias{Dantas2026_TGEXI}]{Dantas2026_TGEXI}, we identified a population of apparently single FGK-type stars at the main-sequence turn-off displaying anomalous UV emission. We used the available spectroscopic, astrometric, and photometric indicators to minimise, as far as possible, contamination from binarity, multiplicity, and blending. To our knowledge, this represented the first explicit attempt to connect a sample of UV-bright, apparently unevolved Galactic-field FGK stars with the extragalactic UV-upturn phenomenon.

These objects are intriguing because stellar activity diagnostics and multiplicity indicators do not provide a uniform explanation for their UV excess. However, H$\alpha$, a key chromospheric activity diagnostic, is unavailable for some stars, particularly in the UV-extreme G1 group, so activity cannot be excluded in every case (see \citetalias{Dantas2026_TGEXI}). They may therefore represent a previously overlooked contribution to the UV output of old stellar populations. Their ages, chemical signatures, Galactic orbits, and formation histories can help determine whether they share a common origin or arise through multiple evolutionary channels.

In this second paper of the series, we extend this connection through, to our knowledge, the first detailed chemical-abundance and chemo-dynamical analysis of a sample such as \tgex. We examine whether the UV-defined groups occupy distinct chemical and orbital regimes, exhibit systematically different dynamical properties, or share detailed enrichment signatures that may provide clues to the origin of their UV emission. This paper is structured as follows. Section~\ref{sec:data_methods} describes the data and methods, Sect.~\ref{sec:results} presents the results and discusses their implications, and Sect.~\ref{sec:conclusions} summarises our conclusions. The data availability is described at the end of this manuscript.

\section{Data and methodology}
\label{sec:data_methods}

\subsection{Data}
\label{subsec:data}

We use the panchromatic data set of  37 main-sequence turn-off stars (MSTO) comprising the \tgex\ catalogue, constructed in \citetalias{Dantas2026_TGEXI}. It combines high-resolution optical spectroscopy from the final public data release of the \gaia-ESO Survey \citep[iDR6, equivalent to DR5;][]{Gilmore2012, Gilmore2022, Randich2013, Randich2022} with \gaia\ DR3 astrometry and photometry \citep{Gaia2023_DR3}, GALEX UV photometry \citep[NUV and FUV;][]{Martin2005, Bianchi2014}, and infrared photometry from 2MASS \citep[$JHK_s$;][]{Skrutskie2006} and AllWISE \citep[$W1W2$;][]{Cutri2014}. Only spectra with S/N~$>40$ were retained, and the sample was cleaned using multiplicity indicators and quality flags from both \gaia-ESO and \gaia. 

All stars in the larger \gex\ parent catalogue ($\sim$ 3000 stars) have the remaining measurements, whereas some lack FUV photometry; \tgex\ is the clean subset with both FUV and NUV data. The stellar identifiers retain the numbering of the parent \gex\ catalogue, which follows the same selection criteria; \tgex\ is its FUV-complete subset. The parent catalogue will be explored separately.

In \citetalias{Dantas2026_TGEXI}, the sample was classified into three groups through the Gaussian mixture model analysis \citep[GMM;][]{McLachlan2000, Hastie2001, Murphy2013}. G1 contains the stars with the strongest UV excess (5 stars), G2 comprises the predominantly UV-normal population (11 stars), and G3 occupies an intermediate regime (21 stars), displaying a milder UV excess than G1. We retain this nomenclature throughout the present paper. 

We adopted the stellar ages and uncertainties derived primarily with \textsc{UniDAM} \citep{Mints2017, Mints2018} using PARSEC isochrones \citep{Bressan2012}. The complete data selection, treatment, and GMM clustering are described in \citetalias{Dantas2026_TGEXI}.

\subsection{Calculation of $\alpha$-enrichment abundances}
\label{subsec:alpha_abundances}

To characterise the overall $\alpha$-element enrichment, we calculated \afe\ for each star using the subset of Mg, Si, Ca, and Ti measurements available for that individual star. The quantity \afe\ was defined as the inverse-variance-weighted mean of the available abundance ratios,

\begin{equation}
\label{eq:alpha_fe}
    [\alpha/\rm{Fe}]
    =
    \frac{\sum_i w_i[\rm{X}_i/\rm{Fe}]}
         {\sum_i w_i},
    \qquad
    w_i = (\sigma_{\rm{X}_i}^{2} + \sigma_{\rm{Fe}}^2)^{-1},
\end{equation}

\noindent where the sum runs over the available measurements of Mg, Si, Ca, and Ti, with $\sigma_{\rm{X}_i}$ and $\sigma_{\rm{Fe}}$ denoting the uncertainties in
$[\rm{X}_i/\rm{H}]$ and \feh, respectively. Because \gaia-ESO does not provide abundance covariances, we treated the uncertainties as independent; hence, the uncertainty of the weighted mean was computed as

\begin{equation}
\label{eq:alpha_fe_uncertainty}
    \sigma_{[\alpha/\rm{Fe}]}^2 =
    \left(\sum_i w_i\right)^{-1}.
\end{equation}

Because abundance measurements are not available for every element in every star, individual \afe\ values may be based on different subsets of the four adopted $\alpha$-elements. In particular, two of the 37 \tgex\ stars lack \mgfe\ measurements. A comparison between the resulting \afe\ distributions and the directly measured \mgfe\ distributions is provided in Figure~\ref{fig:alpha_Mg_violin}.

\subsection{Orbit integration}
\label{subsec:orbits}

We integrated the Galactic orbits of the 37 \tgex\ stars using radial velocities from the \gaia-ESO survey and astrometric parameters from \gaia\ DR3, including positions, parallaxes, proper motions, and their associated uncertainties. \gaia\ parallax zero-point corrections were applied using the recipe of \citet[][and obligatorily using \texttt{pseudocolor}=\texttt{NAN} or between 1.24 and 1.72, and \texttt{astrometric\_params\_solved} $>$3.]{Lindegren2021}; Bayesian distances were estimated via the prescription of \citet{Bailer-Jones2015}. A summary of the resulting distance and spatial distribution of the \tgex\ stars is provided in Figure~\ref{fig:distances}.

The orbits were integrated backwards for 10 Gyr with \textsc{Galpy} \citep{Bovy2015}, adopting the MW potential of \citet{McMillan2017}. We used the Dormand--Prince integration scheme implemented in \textsc{C} (\texttt{method=`dop853\_c'}; \citealt{DormandPrince1980}). The Galactocentric transformation and adopted Solar parameters follow \citet{Dantas2023}. Angle brackets denote medians across the 100 bootstrap realisations of the observed phase-space parameters. For each star, we derived distributions of the Galactic space velocities and relevant orbital quantities, including the guiding radius (\rgui), pericentric and apocentric radii, eccentricity, maximum vertical excursion from the Galactic plane (\zmax), angular momentum ($L$), and orbital actions. Unless otherwise stated, we adopt the median of each distribution as the representative value.

\subsection{Hierarchical clustering}
\label{subsec:hc}

We investigated the chemical structure of \tgex\ using agglomerative hierarchical clustering \citep[HC;][]{Murtagh2014, Murtagh&Contreras2012}. The clustering used Ward's minimum-variance criterion and Euclidean distances in the 13-dimensional space defined by 13 abundance measurements from 11 elements: \ion{Mg}{i}, \ion{Al}{i}, \ion{Si}{i}, \ion{Ca}{i}, \ion{Ca}{ii}, \ion{Sc}{ii}, \ion{Ti}{i}, \ion{Ti}{ii}, \ion{Cr}{i}, \ion{Mn}{i}, Fe\footnote{The Fe abundance was estimated from the \feh\ provided by \gaia-ESO iDR6 and therefore has no associated ionisation level.}, \ion{Co}{i}, and \ion{Ni}{i}. Only stars with all 13 measurements were clustered; missing abundances were not imputed.

Stars with incomplete abundance coverage were placed a posteriori within this ordering according to their similarity in \teff, \logg, age, \fuvnuv, and \nuvg, after min--max scaling to the interval $[0,1]$. This step uses stellar-parameter similarity only to organise the full sample for visual comparison; it neither alters the chemical hierarchy nor defines independent populations.

\section{Results, analysis, and discussion}
\label{sec:results}

\subsection{Group-wise correlations and general features}
\label{sec:group_wise_correlations}

\begin{figure*}[ht!]
    \centering
    \includegraphics[width=\linewidth,trim={5mm 7mm 5mm 7mm},clip]{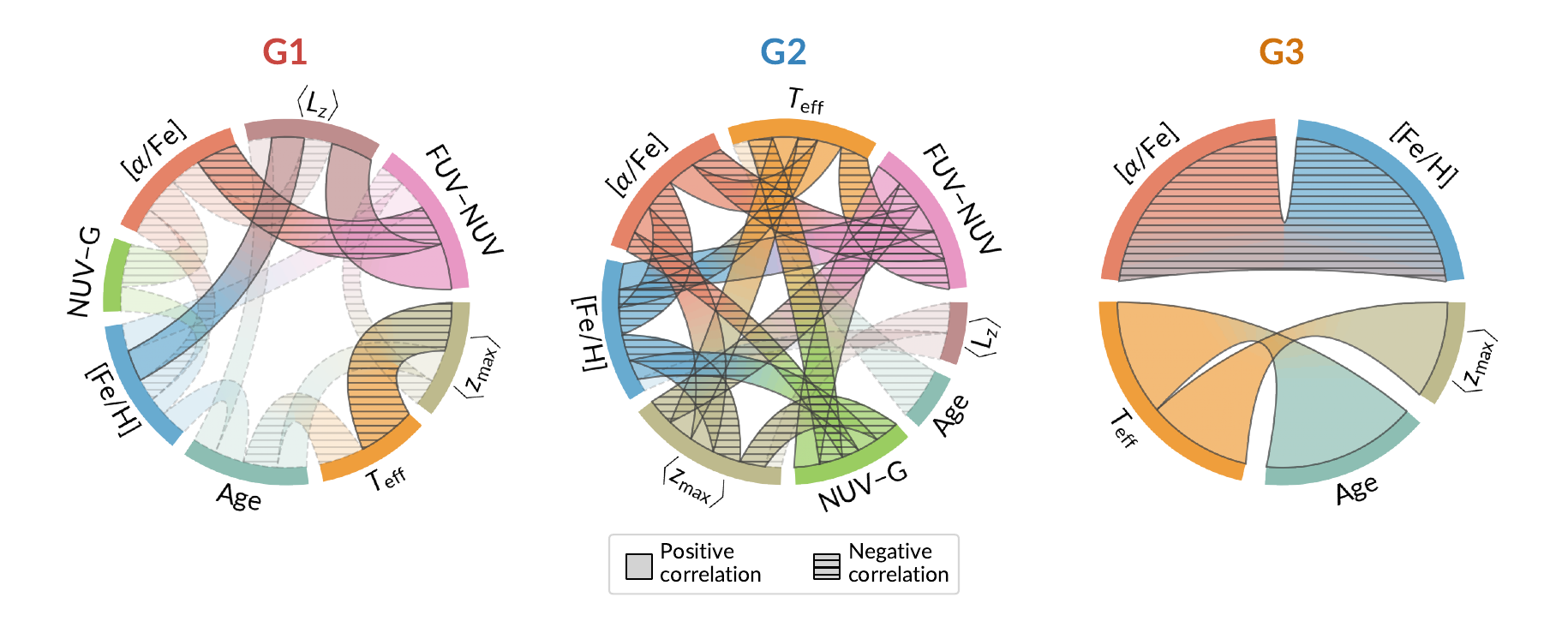}
    \caption{Group-wise chord diagrams of the strongest Spearman correlations among the stellar, chemical, UV, and orbital parameters. Only correlations with $|\rho|\geq0.4$ are displayed. Solid chords have bootstrap--Monte Carlo confidence intervals that remain at $|\rho|\geq0.1$, as explained in the main text; semi-transparent chords have intervals extending below this value. Unhatched and hatched chords denote positive and negative correlations, respectively. Given the small group sizes, the diagrams are descriptive. They were made with \cachai\ \citep{Beltran2025}.}
    \label{fig:group_correlations}
\end{figure*}

Figure~\ref{fig:group_correlations} presents the strongest correlations within each GMM group as chord diagrams constructed with \cachai\ \citep[][but see also \citealt{Gu2014} and \citealt{deSouzaCiardi2015}]{Beltran2025, Beltran2026_MichiRNAAS}. As in Fig. 5 of \citetalias{Dantas2026_TGEXI}, where correlations were computed for the full \tgex\ sample, we calculate Spearman rank correlations \citep[$\rho$,][]{Spearman1904} separately within each group to examine how the global correlation structure changes when the sample is stratified. Here, $\rho$ denotes the Spearman rank correlation coefficient between a pair of parameters\footnote{The $\rho$ coefficient can range from $-1$ to $1$, with $\rho = 1$ and $\rho = -1$ corresponding to perfect positive and negative rank correlations, respectively, and $\rho = 0$ indicating no correlation.}. All correlations are calculated, but \cachai\ displays only those with $|\rho|\geq0.4$, omitting weaker correlations that are not sufficiently strong to warrant inclusion in the chord diagrams.

Given the small and unequal numbers of stars in the groups, these diagrams should therefore be interpreted as a descriptive comparison of the correlation structures within the observed \tgex\ subsamples, rather than as evidence for intrinsic relations in the broader Galactic population. To assess the robustness of the observed correlations in the presence of these small samples, we use bootstrap resampling (5000 realisations) combined with Monte Carlo propagation of the observational uncertainties \citep[following the method described in][]{Curran2014}. The resulting distributions of $\rho$ are used to estimate confidence intervals for each correlation coefficient. Among the displayed correlations, those for which the 16th percentile (consistent with -1$\sigma$) remains at $|\rho| \geq 0.1$ are shown as solid chords, whereas those whose confidence interval extends to $|\rho| < 0.1$ are shown as semi-transparent chords, due to their extremely low values for -1$\sigma$.

The age distributions are broad: G1 and G3 both have median ages of $5.89$~Gyr, with 16th--84th percentile intervals of $0.03$--$11.16$ and $0.02$--$12.30$~Gyr, respectively, whereas G2 has the youngest median, $4.27$~Gyr ($2.39$--$7.71$~Gyr). The formally youngest G1 and G3 solutions are likely affected by isochrone degeneracy, as discussed in \citetalias{Dantas2026_TGEXI}.

\paragraph{Group 1:}
\label{par:g1}

G1, which contains the strongest UV-excess stars, shows selective links between the UV colours and chemical and orbital properties. \fuvnuv\ is linked most strongly to \afe\ and \lz, and additionally to \feh\ and \zmax, whereas \nuvg\ is linked mainly to the chemical abundances. \lz\ also links to \feh, and \zmax\ is anti-correlated with \teff. Thus, the UV-colour variation in G1 covaries with chemical composition and Galactic dynamics, not with \teff\ alone.

\paragraph{Group 2:}
\label{par:g2}

The UV-normal G2 group has the densest network. \teff\ occupies a central position, linking to both UV colours, the chemical abundances (positively with \afe\ and negatively with \feh) and the orbital quantities, consistent with the strong $\teff$--(\fuvnuv) relation in \citetalias{Dantas2026_TGEXI}. The UV colours also link to several non-UV parameters. Part of this greater network density may reflect the group size (11 stars) and parameter range.

\paragraph{Group 3:}
\label{par:g3}

G3 has a much sparser network: no correlation involving the UV(-optical) colours reaches the adopted threshold. The remaining links involve \teff, age, \zmax, \feh, and \afe, including the expected \feh--\afe\ anti-correlation. The moderate positive $\teff$--age correlation should be treated cautiously because some age estimates may be affected by the isochrone degeneracy discussed in \citetalias{Dantas2026_TGEXI}. Thus, no individual stellar, chemical, or orbital parameter is strongly associated with the UV colours at the adopted threshold, although the small group size limits the stability of the coefficients.


\paragraph{Global assessment:}
\label{par:global_assessment}

The three group-wise networks therefore share no common set of UV-colour associations, although the \feh--\afe\ anti-correlation recurs. Only G3 retains an age-related chord under the bootstrap criterion; the age correlations in G1 and G2 do not retain such correlation. The distinct G1 and G3 networks, despite both groups containing UV-excess stars, indicate that the UV-abnormal population is not homogeneous and may reflect different combinations of stellar, chemical, and dynamical properties. These correlations cannot establish distinct physical origins, which should be tested with a larger sample. Fig. \ref{fig:all_michinoff} provides a complementary multivariate view through the use of customised Chernoff Faces visualisation.

\subsection{Galactic components and birth environments}
\label{subsec:dynamics}

\subsubsection{Toomre and Lindblad diagrams}
\label{subsubsec:toomre_lindblad}

\begin{figure}[ht!]
    \centering
    \includegraphics[width=\linewidth,trim={5mm 7mm 5mm 7mm},clip]{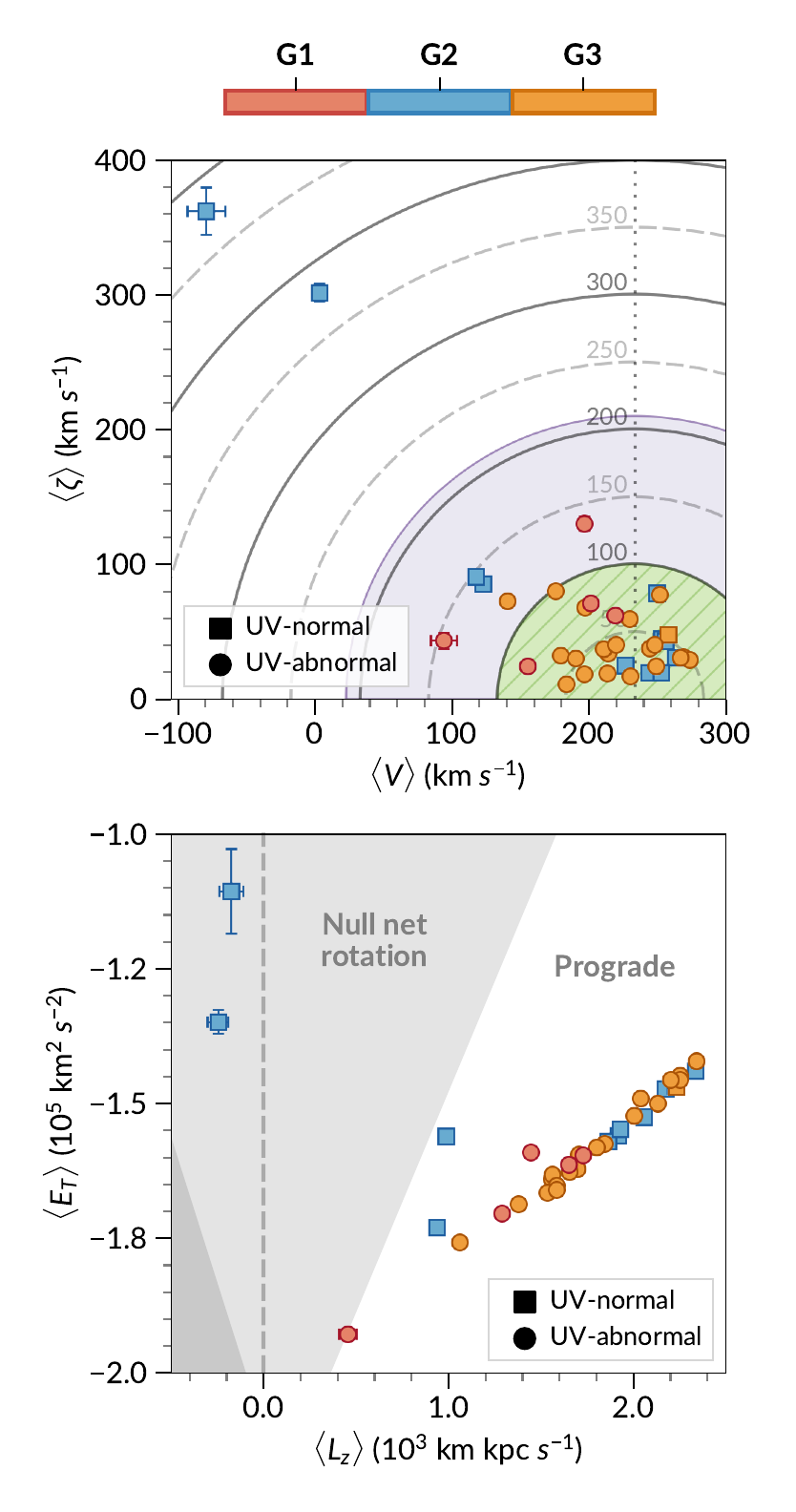}
    \caption{Toomre (top) and Lindblad (bottom) diagrams for the \tgex\ sample. Colours identify the GMM groups, while marker shapes indicate the UV classification. In the Toomre diagram, the velocity contours are centred on $V_{\odot}=232.8 {\rm km~s^{-1}}$ \citep{McMillan2017}. The regions enclosed by the $100~{\rm km\,s^{-1}}$ contour, between the $100$ and $210 {\rm km~s^{-1}}$ contours, and beyond the $210 {\rm km~s^{-1}}$ contour are used here to identify disc-like, disc--halo transition, and halo-like kinematics, respectively \citep{Bensby:2003, Helmi:2018}. In the Lindblad diagram, the dashed line indicates $\langle L_z\rangle=0$, and the shaded areas delimit retrograde, low-net-rotation, and prograde orbital regimes following the empirical divisions adopted by \citet{Giribaldi:2023}.}
    \label{fig:toomre_lindblad}
\end{figure}

Figure~\ref{fig:toomre_lindblad} presents the kinematics and orbital properties of \tgex. The Toomre diagram combines the azimuthal velocity, $\langle V\rangle$, with $\langle\zeta\rangle=\langle\sqrt{U^2+W^2}\rangle$, calculated in each orbit realisation before adopting the median. The Lindblad diagram relates the total binding energy, $\langle E_T\rangle$, to \lz. The two projections respectively highlight departures from the Solar velocity and prograde, retrograde, or low-angular-momentum configurations.

Thirty of the 37 stars lie within the $100~{\rm km~s^{-1}}$ thin disc contour, five occupy the intermediate-velocity region between the two contours (commonly associated with a kinematically defined thick disc and referred to here as the transition region) and two lie beyond the $210~{\rm km~s^{-1}}$ halo contour. G1 contains three disc-like and two transition-region stars. G2 spans the widest range, with two transition-region and two halo-like objects in addition to its disc-like majority. All but one G3 star lie within the thin disc contour.

All G1 and G3 stars are prograde, whereas G2 contains the only two low-net-rotation objects, which are also the two halo-like stars in the Toomre diagram. They need not share an origin: GEX0173 (\feh$=-1.35$) lies within the metallicity range commonly associated with \gaia-Enceladus-Sausage, whereas GEX1383 (\feh$=-0.27$) is compatible with the Splash population. The lack of an $\alpha$-abundance measurement for GEX1383 prevents a firmer classification \citep[see e.g.][]{Giribaldi:2023}. The remaining stars occupy a comparatively narrow prograde distribution, apart from several objects with thick-disc-like velocities.

Thus, the UV excess is not preferentially associated with halo-like or non-rotating populations: G1 and G3 are confined to prograde orbits and are predominantly disc-like. This differs from the UV excesses reported among very metal-poor halo stars \citep[such as in][]{Smith:2026} and may indicate different physical physical mechanisms operating across Galactic populations.

\subsubsection{The Tinsley--Wallerstein diagram}
\label{subsubsec:tw_diagram}

\begin{figure}[ht!]
    \centering
    \includegraphics[width=\linewidth,trim={5mm 7mm 5mm 7mm},clip]{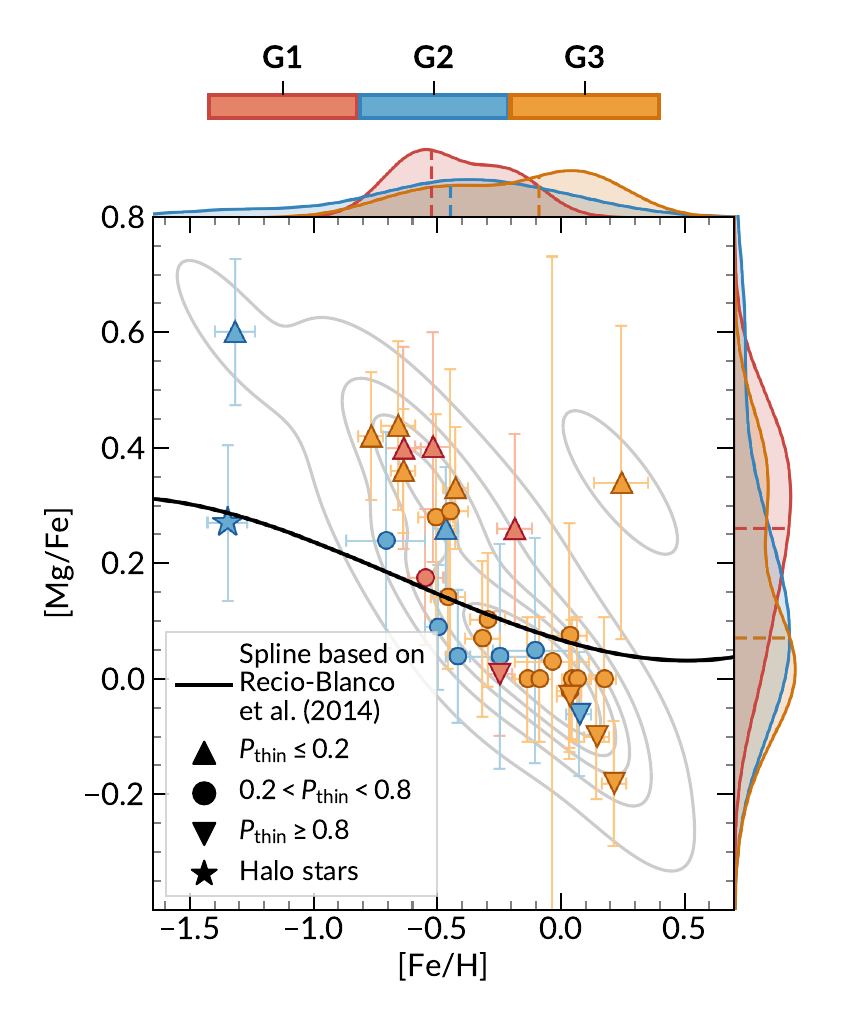}
    \caption{Tinsley--Wallerstein diagram for the \tgex\ sample. Colours identify the GMM groups. The solid curve represents the empirical thin/thick-disc boundary obtained by combining the relations of \citet{RecioBlanco2014} with extensions to larger range of values using \citet{Adibekyan:2012} and interpolating them with a cubic spline. Upward triangles, circles, and downward triangles identify stars with $P_{\rm thin} \leq 0.2$, $0.2 < P_{\rm thin} < 0.8$, and $P_{\rm thin} \geq 0.8$, respectively. Star-shaped markers denote stars classified as belonging to the halo.}
    \label{fig:tinsley_wal}
\end{figure}

Figure~\ref{fig:tinsley_wal} presents the Tinsley--Wallerstein diagram \citep{Wallerstein1962, Tinsley1979} for the \tgex\ sample. The three groups occupy broadly overlapping regions of the \mgfe--\feh\ plane and do not define separate chemical sequences. Nevertheless, their distributions show systematic differences: G1 extends towards lower \feh\ and higher \mgfe, whereas G2 and G3 are more evenly distributed across the metal-poor and metal-rich regimes. The diagram therefore provides a 2D view of the group-level metallicity and $\alpha$-enhancement differences reported in Fig.~3 of \citetalias{Dantas2026_TGEXI}.

The adopted boundary for the chemical separation between thin and thick discs follows \citet{RecioBlanco2014}, with the metal-poor and metal-rich extensions of \citet{Adibekyan:2012}; the piecewise thresholds were joined by cubic-spline interpolation. For each star, we drew $10^{5}$ Monte Carlo realisations with Gaussian errors in \feh\ and \mgfe. The fractions below and above the boundary define $P_{\rm thin}$ and $P_{\rm thick}\equiv1-P_{\rm thin}$, respectively. As shown in Figure~\ref{fig:tinsley_wal}, we adopt $P_{\rm thin}\leq0.2$, $0.2<P_{\rm thin}<0.8$, and $P_{\rm thin}\geq0.8$ as conservative criteria for thick-disc-compatible, chemically ambiguous, and thin-disc-compatible stars, respectively, given the substantial propagated uncertainties in \mgfe\ for several stars. The large uncertainties of several stars prevent an unambiguous classification.

G1 is the most $\alpha$-enhanced group, with systematically higher \mgfe; its \afe\ distribution is also more sharply concentrated around its median (Appendix \ref{append_sec:mgfe_vs_afe}). Most G1 stars lie above the boundary at their central values. Three of the five satisfy $P_{\rm thin}\leq0.2$, one is ambiguous, and one is thin-disc-compatible; thus, all but the latter remain compatible with thick-disc membership to varying degrees. G2 is predominantly thin-disc-compatible, although at least two stars favour the thick disc, while G3 shows no preference. The principal distinction is therefore G1's collective shift towards higher \mgfe, not an exclusive correspondence between a GMM group and a Galactic disc component.

Two objects depart markedly from the main distribution through their high \mgfe. Both are comparatively hot (\teff~$\sim6500$--$7000$~K) and young ($\sim1$~Gyr), but otherwise differ substantially: one is metal-poor and UV-normal, while the other is metal-rich and UV-abnormal. More generally, the UV-abnormal stars do not occupy a unique locus in the Tinsley--Wallerstein plane beyond the group-level metallicity and $\alpha$-enhancement trends described above.

G1's collective $\alpha$-enhancement has a qualitative analogue in unresolved galaxies, where UV-upturn strength correlates with Mg$_2$ or Mg~$b$ and with \afe\ \citep{Burstein1988, Donas2007, Bureau2011, Carter2011}. All five G1 stars also have high absolute \mgh, and their measured [\ion{Ti}{i}/Fe] and [\ion{Ti}{ii}/Fe] ratios lie above zero, showing that the group-level enhancement is not confined to Mg. However, integrated Mg-sensitive indices trace the optically dominant population and do not establish that the UV-emitting stars themselves are Mg-enhanced; the resolved and unresolved signals may instead reflect complementary signatures of a common enrichment history.

\subsubsection{Birth radii of probable thin disc members}
\label{subsubsec:birth_guiding_radii}

To investigate whether the \tgex\ groups trace different Galactic formation regions, we estimated their birth radii, \rb. The methodology adopted here was developed in \citet{Dantas2025a} and combines a generalised additive model \citep[GAM;][]{HastieTibshirani1990} with the Galactic thin-disc chemical-evolution models of \citet{Magrini2009}\footnote{Related semi-empirical approaches using stellar age and \feh\ to estimate stellar \rb\ have been explored and developed in several studies \citep[e.g.][]{Minchev2018, Ratcliffe2023, Lu2024, Ratcliffe2024a, Ratcliffe2024b, Ratcliffe2025}. We refer to \citet{Dantas2025a} for a detailed description of the methodology adopted here, including its validation and limitations.}. We compared the resulting \rb\ among the GMM groups and with the present-day guiding radii, \rgui, obtained from the orbit integrations.

Because the \citet{Magrini2009} models describe the thin disc, we restricted the analysis to the 18 stars satisfying (1 in G1, 5 in G2, and 12 in G3) both the disc kinematic classification in the Toomre diagram and a chemical thin disc probability of $P_{\rm thin}>0.5$, as defined above. Marker shapes distinguish stars with $P_{\rm thin}\geq0.8$ from those with $0.5<P_{\rm thin}<0.8$.

\begin{figure}
    \centering
    \includegraphics[width=\linewidth,trim={7mm 7mm 5mm 7mm},clip]{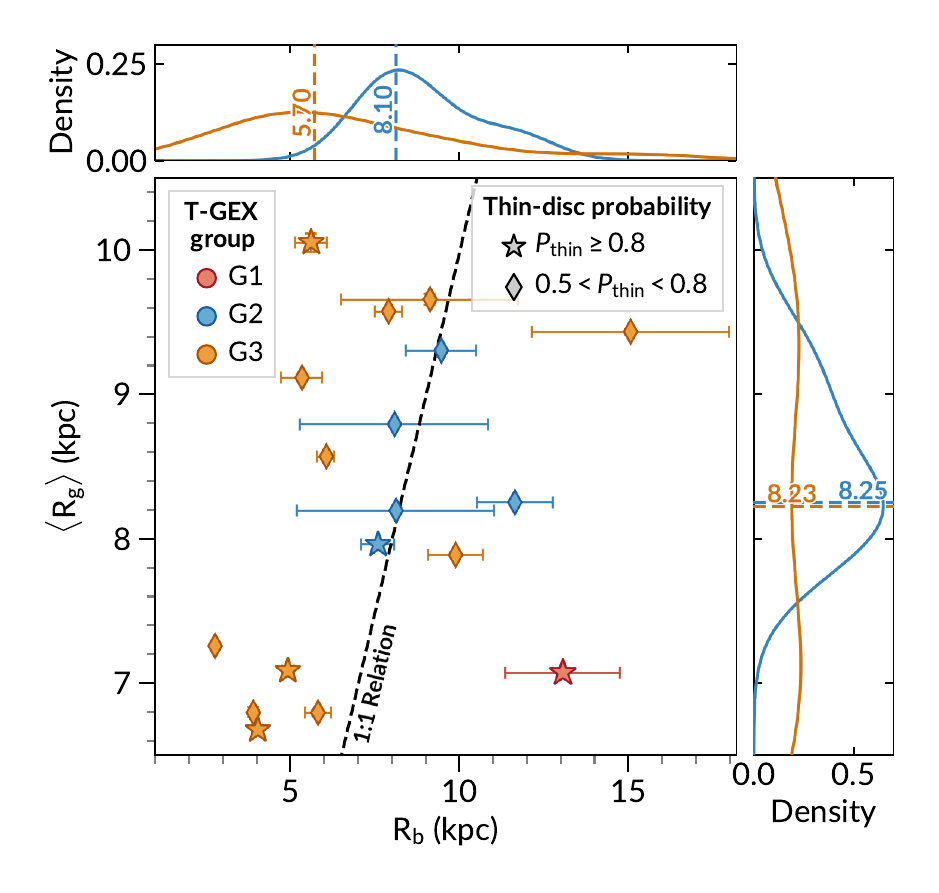}
    \caption{Present-day \rgui\ versus inferred birth radii, \rb, for the probable thin-disc members of \tgex. Colours identify the GMM groups, while star- and diamond-shaped markers denote stars with $P_{\rm thin}\geq0.8$ and $0.5<P_{\rm thin}<0.8$, respectively. Horizontal error bars show formal GAM prediction uncertainties, excluding uncertainties in the input ages and \feh\ and chemical-evolution-model systematics. The dashed diagonal marks \rgui=\rb. The marginal KDEs show the independently normalised \rb\ and \rgui\ distributions of G2 and G3, with their medians indicated by dashed lines. G1 is omitted from the KDEs because only one G1 star satisfies the combined selection.}
    \label{fig:guiding_and_birth_radii}
\end{figure}

The principal group-level difference occurs in \rb. The intermediate-UV G3 stars have a $\overline{R_{\rm b}}$ of $5.70$~kpc, compared with $8.10$~kpc for the UV-normal G2 stars, whereas their $\overline{\langle R_{\rm g} \rangle}$ are nearly identical at $8.23$ and $8.25$~kpc, respectively. Thus, within the selected sample, G3 is inferred to have formed farther inward than G2 despite the groups currently occupying the solar vicinity. G3 retains the smaller median \rb\ in both chemical thin-disc probability intervals examined in Appendix~\ref{append_sec:supplement} through Figure~\ref{fig:rbirth_rgui_kde}.

The G1 group is dominated by thick-disc objects with only one star, GEX1170, satisfying the combined thin-disc selection; its inferred \rb\ is larger than its \rgui\ and cannot characterise G1 as a group. Because the adopted chemical evolution mapping generally assigns smaller \rb\ at fixed \feh\ to older ages, an underestimated age for GEX1170 would shift its inferred \rb\ inward, potentially towards the values found for G3. This remains conditional on the uncertain age of a single star and cannot establish a group-level G1 trend.

\subsection{Detailed chemical signatures and implications}
\label{subsec:chemical_abund}

\subsubsection{Individual chemical abundances}
\label{subsubsec:individual_abund}

\begin{figure}
    \centering
    \includegraphics[width=\linewidth,trim={1.4cm 7mm 2.3cm 7mm},clip]{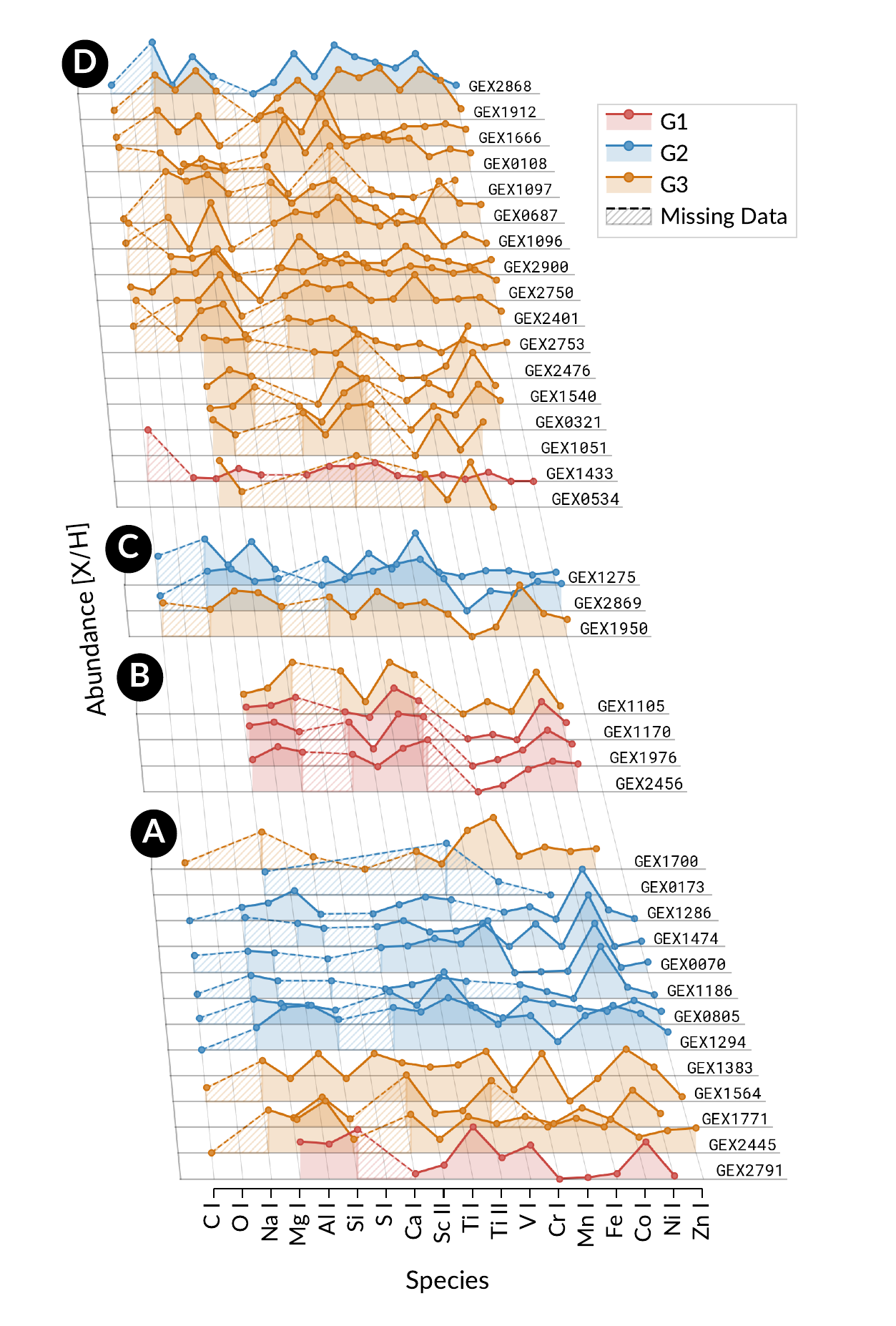}
    \caption{Stacked chemical-abundance patterns ([X/H]) for the 37 \tgex\ stars shown in a compact representation. The sample is arranged from sections A--D following the HC chemical ordering and, within each panel, by increasing \teff. Colours identify the GMM groups defined in \citetalias{Dantas2026_TGEXI}, similarly to previous figures. Solid segments connect adjacent abundance measurements, while dashed segments indicate species with no reported abundance by \gaia-ESO. Identifiers are indicated for each pattern. The detailed version of this figure, including individual $1\sigma$ uncertainties, is provided in Figure~\ref{fig:chemical_patterns}.}
    \label{fig:chemical_patterns_simplified}
\end{figure}

Figure~\ref{fig:chemical_patterns_simplified} compares the individual [X/H] profiles using the HC-informed ordering described in Sect.~\ref{subsec:hc}. For stars with complete abundance coverage, this ordering is derived from the HC based on all 13 measurements; stars with incomplete coverage were placed a posteriori, and sections A--D divide the sequence for display rather than define independent populations. The profiles show substantial diversity, and the HC-informed sections contain different proportions of the UV-defined populations. Most notably, Group B contains 3 of the 5 strongly UV-excess G1 stars and no UV-normal G2 objects. Its G1 members are comparatively cool, with \teff\ between approximately 4800 and 5000~K.

The most distinctive recurring feature of Group B occurs among the Fe-peak elements \ion{Cr}{i}, \ion{Mn}{i}, \ion{Fe}{i}, \ion{Co}{i}, and \ion{Ni}{i}. In GEX2456 and GEX1976, Cr and Mn lie below Fe, the pattern rises towards Co, which lies above Fe, and then declines towards Ni. Although the profiles are displayed as [X/H], the positions of these elements relative to Fe correspond to comparatively low [\ion{Cr}{i}/Fe] and [\ion{Mn}{i}/Fe] together with enhanced [\ion{Co}{i}/Fe]. GEX1170 and the intermediate-UV star GEX1105 also show enhanced Co followed by a decrease towards Ni, although their progression from Cr to Co is less regular.

Comparable Ti--Fe-peak chemical-abundance sequences occur in all five G1 stars, despite their placement in three different HC display sections. All three G1 stars in panel B show this morphology, while GEX2791 and GEX1433 show similar sequences in panels A and D, respectively. Their different placements reflect the complete abundance profiles used in the HC analysis rather than this restricted set of elements. For GEX1433, the unusually high \ion{C}{i} abundance expands the vertical range and visually compresses the remaining variations. When examined on their own scale, its abundances show the same broad progression from enhanced \ion{Ti}{i} and \ion{Ti}{ii}, through the Cr--Mn--Fe region towards Co, followed by a decline through Ni and the uniquely measured Zn.

The Co-rich G2 stars GEX1186, GEX1474, and GEX1286 are among the hottest objects in the sample, with \teff\ between approximately 6580 and 6830~K, in contrast to the cool G1 stars concentrated in panel B. They are also distinct from GEX0173 and GEX1383, the two G2 objects with halo-like velocities and low angular momenta. Their prominent Co abundances are therefore not associated simply with the dynamically distinct G2 population, while their common \teff\ regime raises the possibility of additional \teff-dependent abundance systematics.

\begin{figure*}
    \centering
    \includegraphics[width=\linewidth]{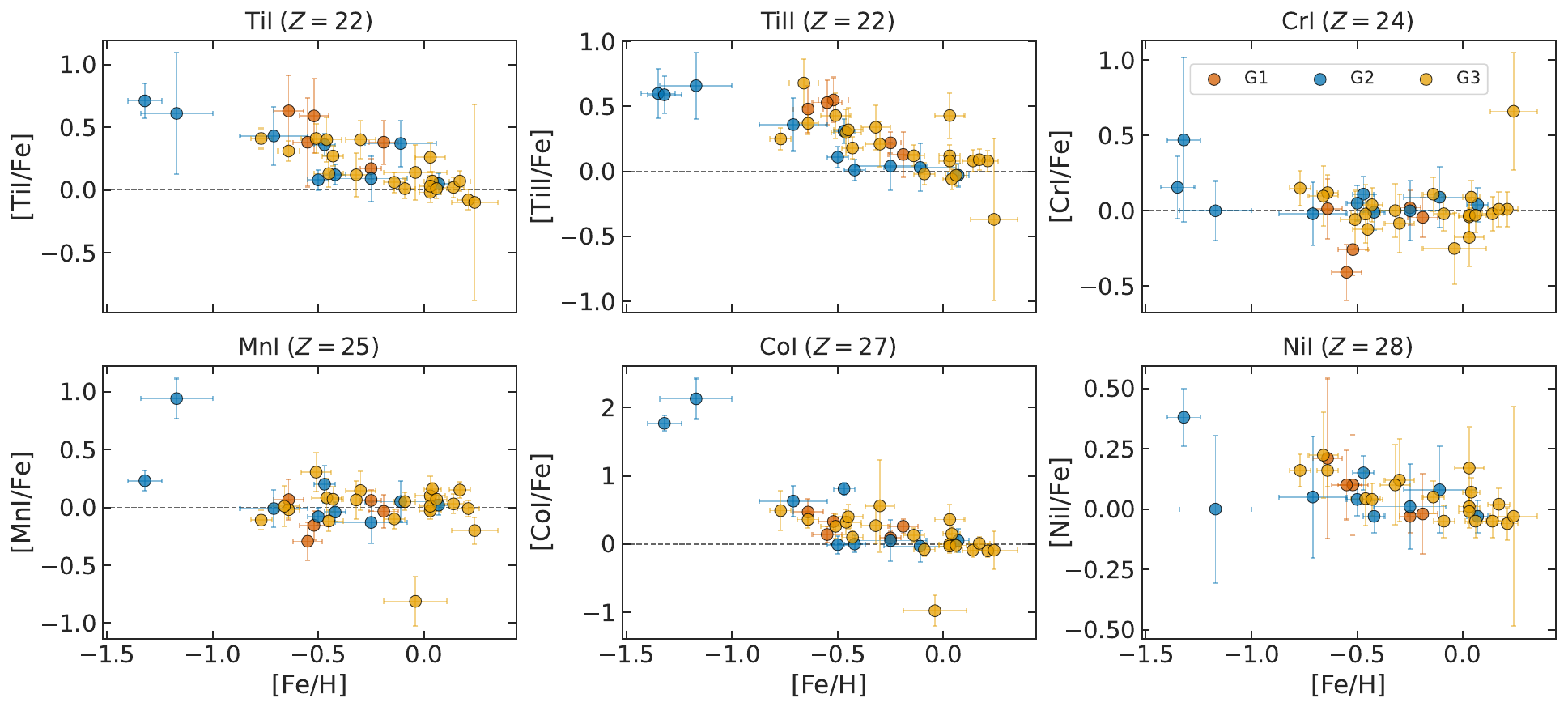}
    \caption{Titanium and Fe-peak abundance ratios as a function of \feh\ for the \tgex\ stars with available measurements. The panels show [\ion{Ti}{i}/Fe], [\ion{Ti}{ii}/Fe], [\ion{Cr}{i}/Fe], [\ion{Mn}{i}/Fe], [\ion{Co}{i}/Fe], and [\ion{Ni}{i}/Fe], with the corresponding atomic number, $Z$, indicated in each title. All ratios were calculated as $[\mathrm{X/Fe}]=[\mathrm{X/H}]-[\mathrm{Fe/H}]$ using the adopted survey \feh. Colours identify the GMM groups defined in \citetalias{Dantas2026_TGEXI}: G1 contains the stars with the strongest UV excess, G2 the UV-normal stars, and G3 the intermediate-UV population. Error bars represent the propagated $1\sigma$ uncertainties.}
    \label{fig:ti_iron_peak_ratios}
\end{figure*}

Figure~\ref{fig:ti_iron_peak_ratios} places these abundance contrasts in the context of the metallicity distribution of the sample. Both Ti ionisation stages indicate that all five G1 stars are Ti-enhanced relative to Fe, with G1 tending to occupy the upper portion of both Ti distributions over its metallicity range. The agreement between the two ionisation stages lends additional robustness to this enhancement and supports the broader $\alpha$-enhancement.

Viewed jointly, the abundance ratios reveal a structured group-level result. All five G1 stars have positive central values of [\ion{Ti}{i}/Fe], [\ion{Ti}{ii}/Fe], and [\ion{Co}{i}/Fe], while Cr and Mn generally occupy lower ratios and are subsolar in several objects. The amplitudes vary from star to star, but the same overall Ti--Cr--Mn--Co direction recurs throughout G1. This repeated co-occurrence is not reproduced by the isolated Co peaks among the UV-normal G2 stars. Its appearance in a related form in some G3 stars further suggests that the signature may extend across the UV-excess population, with its clearest systematic recurrence in G1. Although the broader metallicity dependence of Mn may contribute to the amplitude of one component, it does not by itself account for the convergence of Ti, Cr, Mn, and Co within the same objects.

\subsubsection{A hypernova-like chemical morphology}
\label{subsubsec:hypernovae_morphology}

The combination of enhanced Ti and Co with comparatively low Cr and Mn closely resembles the abundance signature predicted for high-energy core-collapse supernovae, or hypernovae. The elements Cr and Mn are produced predominantly through incomplete Si burning, whereas Fe and Co primarily trace complete Si burning. Because Fe and Co originate from closely related burning conditions, enhanced [Co/Fe] represents a change in their relative yields rather than a general enhancement of the Fe-peak elements. In higher-energy explosions, the complete Si-burning region extends over a larger fraction of the ejecta relative to the incomplete Si-burning region, producing higher [Co/Fe] together with lower [Cr/Fe] and [Mn/Fe]. The accompanying $\alpha$-rich freeze-out enhances the production of $^{48}$Cr, which subsequently decays through $^{48}$V into $^{48}$Ti, while also favouring the production of Zn. Hypernova nucleosynthesis is therefore characterised by larger Ti, V, Co, and Zn abundances relative to Fe and smaller Mn and Cr abundances relative to Fe than normal core-collapse supernovae \citep[see e.g.][]{Umeda2002, Nomoto2006, Nomoto2013, Grimmett2020}.

Observational precedents support the physical plausibility of such localised enrichment. The exceptional Be abundance of the halo dwarf HD~106038, together with its enhanced Li, Si, and Ni, led \citet{Smiljanic2008} to propose that the star formed from natal material enriched by a nearby hypernova. More recently, \citet{Molaro2026} showed that HD~106038, HD~132475, and the other currently known Be-rich stars are associated with the accreted Thamnos-2 structure. Their common Be enhancement, the ratio between the Be and Li excesses, elevated Si abundances, and anomalous $\alpha$-element sequence were interpreted as evidence for a rare energetic enrichment event, potentially a hypernova, followed by different degrees of dilution. In particular, these authors proposed that the same event responsible for the Be production could account for the unusual $\alpha$-enhancement of the population. The \tgex\ stars provide a complementary signature through their Ti and Fe-peak abundances.

Zn, an important additional diagnostic of this enrichment channel, is reported only for GEX1433. Its low absolute \ion{Zn}{i} abundance extends the terminal decline observed in its [X/H] profile, but the relevant hypernova diagnostic is [\ion{Zn}{i}/Fe], since high-energy explosion models predict Zn enhancement relative to Fe. Additionally, abundances derived from neutral Fe-peak lines may be affected by departures from LTE. Published calculations generally yield positive non-local thermal equilibrium (NLTE) abundance corrections for \ion{Cr}{i}, \ion{Mn}{i}, and \ion{Co}{i} \citep{Bergemann2008, Bergemann2010, BergemannPickering2010}. Such corrections could strengthen the inferred Co enhancement, but would also raise the Cr and Mn abundances, thereby weakening their apparent depressions relative to Fe; the net effect on [X/Fe] additionally depends on the corresponding correction to \ion{Fe}{i}. The \ion{Mn}{i} and \ion{Co}{i} measurements are also sensitive to hyperfine splitting. These systematic effects may therefore alter the amplitude of the individual abundance contrasts, but cannot be assumed to strengthen the complete pattern uniformly.

\subsubsection{Chemical composition and the origin of the UV excess}
\label{subsubsec:chem_origin_uv}

Metal-line blanketing may contribute to the emergent UV flux, but in these stars the effect is likely driven mainly by Fe rather than by the detailed Ti--Cr--Mn--Co variations. Fe supplies a dense forest of UV transitions, whereas these less abundant species are more likely to modify individual spectral features than the broadband flux \citep[e.g.][]{Short2005}. The lower \feh\ distribution of G1 could therefore reduce blanketing to some degree. However, the overlap of all three groups near \feh~$\sim-0.25$ and \mgfe~$\sim+0.05$ shows that these abundances alone do not determine the UV classification, and the data do not establish that blanketing can reproduce the full G1 excess. The strong NUV--\feh\ anti-correlation found in much more metal-poor solar-type halo stars supports the general mechanism \citep{Smith:2026}, but tailored spectra are needed to quantify its contribution at the moderate metallicities of G1.

Although an unresolved compact companion cannot be excluded in every system, the multiplicity, blending, and spectroscopic screening in \citetalias{Dantas2026_TGEXI} substantially reduces the likelihood that unrecognised binarity or composite spectra drive the recurring group-level morphology. Mass transfer from an AGB progenitor would more naturally enhance C and $s$-process elements than produce the observed enhanced Ti and Co with comparatively low Cr and Mn \citep[e.g.][]{KarakasLugaro2016, Escorza2019, Kobayashi2020}. GEX1433's high C warrants caution, but without $s$-process measurements (e.g. Sr, Y, Zr, Ba, La, Ce, and Pb) or evidence for a compact companion, it does not establish this channel. Continuum dilution, whereby optical light from an unresolved companion makes the primary star's absorption lines appear weaker, could bias individual abundances but would not naturally reproduce the recurring element-to-element pattern. We therefore favour a natal origin for the recurring chemical abundance pattern, while its connection to the UV excess remains undetermined.

Star--planet interactions provide a possible activity channel. \citet{Shkolnik2013} found tentative evidence for stronger FUV activity in some FGK stars with close-in planets, but no clear dependence on orbital separation or planetary mass and substantial selection effects. After controlling for these biases, \citet{France2018} found no statistically significant contribution. Planetary systems are therefore worth investigating in \tgex, but no host information or evidence currently supports them as a general explanation.

\subsubsection{Implications for unresolved UV-upturn-related stellar populations}
\label{subsubsec:implications}

The association of the UV upturn with massive, metal-rich galaxies \citep[e.g.][and references therein]{Burstein1988, Bureau2011, Dantas2020, Dantas2021} does not require the UV-dominant stars to share the Fe abundance of the optically dominant population. A relatively Fe-poor, $\alpha$-rich minority could contribute disproportionately to the UV while global metallicity and Mg-sensitive indices remain dominated by the broader enriched population. G1 provides a resolved analogue: its UV-extreme stars have a lower and narrower \feh\ range, high absolute \mgh, elevated \mgfe, and an $\alpha$--Fe-peak morphology consistent with rapid massive-star enrichment. G1's lower Fe content could reduce UV line blanketing and strengthen the emergent flux \citep[e.g.][]{Short2005}, but variations in the much less abundant Ti, Cr, Mn, and Co are unlikely to affect the opacity enough to explain the full UV excess. Galaxy-scale UV--metallicity relations may therefore reflect abundance mixture and multiple stellar channels rather than overall metallicity alone.

The \rb\ analysis provides a complementary spatial connection. In spiral galaxies, the old-population UV upturn is generally associated with their bulges \citep{Dorman1995, Oconnell1999}. The probable thin-disc G3 stars have smaller inferred \rb\ than G2 despite nearly identical present-day \rgui, showing that a locally observed intermediate-UV population can originate substantially farther inward than its UV-normal comparison and are consistent with outward migration to the solar region. This is consistent with related stars inhabiting inner-Galaxy environments, including populations contributing to bulge light. It is not direct evidence of bulge membership: the median \rb\ lies in the inner disc, and the GAM's chemical-evolution models do not describe the bulge. Rather, G1 provides the clearest chemical parallel through its $\alpha$-enhancement, whereas G3 provides a tentative spatial parallel through its smaller \rb. Their occurrence in different UV-excess groups supports multiple solar-like stellar populations or evolutionary pathways.

An intriguing, although necessarily speculative, possibility is that the maximum in the UV upturn fraction in red-sequence galaxies around $z \sim 0.25$ \citep{Dantas2020} partly reflects the delayed superposition of these channels in populations formed during or shortly after the peak of cosmic star formation at $z\sim2$ \citep{MadauDickinson2014}. The elevated cosmic star-formation-rate density at that epoch should also have produced a correspondingly higher core-collapse rate \citep{Horiuchi2011, Strolger2015}, increasing the opportunities for rarer high-energy explosions, including hypernovae, to enrich the gas from which subsequent stellar generations formed \citep{Kobayashi2006, Grimmett2020}. By $z\sim0.25$, the resulting low-mass stars would be approximately 7~Gyr old, broadly comparable to the characteristic ages of G1 reported in \citetalias{Dantas2026_TGEXI}, while coeval populations could also supply UV-bright post-main-sequence stars. Establishing whether this combination contributes measurably to the redshift evolution of the UV upturn will require population-synthesis models that jointly follow the canonical evolved channels and the formation and subsequent UV evolution of low-mass stars born from hypernova-enriched gas.

The chemical context also bears on the age-degenerate youngest estimates of G1 reported in \citetalias{Dantas2026_TGEXI}. G1's lower \feh, collective $\alpha$-enhancement, and hypernova-like chemical abundances pattern are more naturally associated with rapid enrichment at earlier, lower-metallicity epochs than with a genuinely very young population, favouring the older isochrone solutions. This is consistent with the larger hypernova contribution expected at early epochs \citep[e.g.][]{Grimmett2020} and the generally older thick disc, despite its age overlap with the thin disc \citep[see e.g.][]{Haywood2013, Bensby2014}. These properties most likely trace natal populations rather than stellar ages and therefore do not provide an independent age determination.

\section{Summary and conclusions}
\label{sec:conclusions}

We present, to our knowledge, the first dedicated chrono-chemo-dynamical investigation of UV-enhanced, apparently single FGK stars in the context of the UV upturn. Our analysis of the 37 \tgex\ stars retains the 3 UV-defined GMM groups identified in \citetalias{Dantas2026_TGEXI} and combines group-wise correlations, Galactic orbit integrations, Toomre and Lindblad diagrams, the Tinsley--Wallerstein plane, and hierarchical clustering of individual chemical-abundance patterns. Given the small sample, the following conclusions apply specifically to \tgex\ and should be interpreted with caution:

\begin{enumerate}

    \item The correlation patterns among stellar properties differ across the three UV-defined groups. In G1, the UV colours covary with \feh, \afe, and orbital quantities. The UV-normal G2 has the densest network, with \teff\ occupying a central position, although this may partly reflect its larger size and broader parameter range. In G3, no correlation involving the UV colours reaches the adopted threshold of $|\rho|\geq0.4$. The contrasting G1 and G3 networks, despite both containing UV-excess stars, indicate that the UV-abnormal population within \tgex\ is not homogeneous.

    \item The \tgex\ sample is dominated by disc-like kinematics: 30 of the 37 stars occupy thin the disc region of the Toomre diagram, 5 lie in the thick-disc--halo region, and only two display halo-like velocities. All G1 and G3 stars follow prograde orbits, while the two halo-like, low-angular-momentum objects belong to the UV-normal G2 population. We therefore find no evidence that the UV excess is preferentially associated with halo-like, retrograde, or low-net-rotation populations.

    \item The three GMM groups overlap substantially in the Tinsley--Wallerstein plane and do not define separate chemical sequences. Nevertheless, G1 is shifted towards lower \feh\ and higher \mgfe, making it the most $\alpha$-enhanced group. All but one of its stars are chemically compatible with thick-disc membership. Its high \mgh\ and collective $\alpha$-enhancement also qualitatively echo the observed relations between UV-upturn strength, Mg-sensitive indices, and \afe\ in unresolved galaxies \citep{Burstein1988, Donas2007, Bureau2011, Carter2011}.

    \item Among the 18 stars satisfying both the disc-like kinematic classification in the Toomre diagram and a chemical thin-disc probability of $P_{\rm thin}>0.5$, G3 has a smaller median inferred \rb\ than G2 ($5.70$ versus $8.10$~kpc), despite their nearly identical median \rgui\ ($8.23$ versus $8.25$~kpc). This ordering persists in both probability subsets, $P_{\rm thin}\geq0.8$ and $0.5<P_{\rm thin}<0.8$, suggesting that G3 preferentially traces more inward formation regions. This provides a possible spatial connection to the bulge-associated UV upturn in spiral galaxies.

    \item The detailed abundance profiles reveal a recurring multi-element morphology among the strongest UV-excess stars. All five G1 stars have positive central [\ion{Ti}{i}/Fe] and [\ion{Ti}{ii}/Fe], while 4 show, with varying prominence, comparatively low Cr and Mn relative to Fe, enhanced Co, and a subsequent decline towards Ni. Three occur consecutively in HC-informed panel B, which contains no G2 stars; individual components outside G1 do not consistently reproduce the full combination.

    \item The convergence of enhanced Ti in both ionisation stages with the recurring Cr--Mn--Fe--Co--Ni morphology is qualitatively consistent with predictions for high-energy core-collapse enrichment. This supports the hypothesis that such events contributed to the natal material of several G1 stars. Zn measurements and a homogeneous treatment will be required to test this interpretation more decisively. Tailored synthetic spectra are needed to quantify any contribution from reduced Fe-line blanketing to the UV excess.
    
    \item Although an unresolved WD companion could contribute to the UV emission of individual systems, mass transfer from its AGB progenitor would not naturally reproduce the observed Ti--Cr--Mn--Fe--Co morphology, which instead favours natal enrichment by high-energy core-collapse events.

\end{enumerate}

In sum, these results do not associate the UV excess with a single Galactic component or chemo-dynamical history. G1 provides the clearest chemical signatures through its collective $\alpha$-enhancement and Fe-peak morphology consistent with hypernova enrichment. Most G1 stars are chemically compatible with the thick disc, for which we cannot provide reliable \rb\ estimates. By contrast, G3 thin disc members preferentially traces more inward formation regions. These complementary chemical and spatial correspondences, clearest in different UV-excess groups, may represent distinct resolved aspects of the composite solar-like stellar populations contributing to the UV upturn and support multiple pathways to anomalous UV emission in apparently unevolved FGK stars. We expect initiatives providing precise chemical abundances for large stellar samples, such as \texttt{CHESS} \citep[see][]{MartinezFernandez2025, Smiljanic2026_IAUS395}, to aid in identifying more such stars.

\section*{Data availability}
The catalogue is only available at the CDS via anonymous ftp to \url{cdsarc.cds.unistra.fr} (130.79.128.5) or via \url{https://cdsarc.cds.unistra.fr/viz-bin/cat/J/A+A/}. Additional parameters can be provided upon reasonable request; please contact the corresponding author for such inquiries.

\begin{acknowledgements}
We thank Melinda Soares-Furtado for the kind suggestions that helped enrich the discussion in this manuscript.
DB, MLLD, and PBT acknowledge ANID Basal Project FB210003. 
MLLD acknowledges Agencia Nacional de Investigación y Desarrollo (ANID), Chile, Fondecyt Postdoctorado Folio 3240344. MLLD is grateful for Miúcha, her feline companion, for the long-standing companionship, love, and support.
DB and MLLD acknowledge support from the Pontificia Universidad Católica de Chile for the summer program ``IPRE''. 
RS acknowledges support from the National Science Centre, Poland, project 2019/34/E/ST9/00133.
PBT acknowledges partial support from Fondecyt Regular 1240465.
This work benefited from the following online platforms: \texttt{slack} (\url{https://slack.com/}), \texttt{github} (\url{https://github.com/}), and \texttt{overleaf} (\url{https://www.overleaf.com/}). Additionally, this work used the following \textsc{python} packages: \textsc{matplotlib} \citep{Hunter2007}, \textsc{numpy} \citep{Harris2020}, \textsc{pandas} \citep{McKinney2010}, \textsc{seaborn} \citep{Waskom2021}, and \textsc{astropy} \citep{Astropy2013, Astropy2018, Astropy2022}.
\end{acknowledgements}

\bibliographystyle{bibtex/aa}          
\bibliography{paper.bib}               

\appendix
\onecolumn

\section{Comparison between \mgfe\ and \afe}
\label{append_sec:mgfe_vs_afe}

\begin{figure*}[ht!]
    \centering
    \includegraphics[width=0.8\linewidth,trim={5mm 7mm 5mm 7mm},clip]{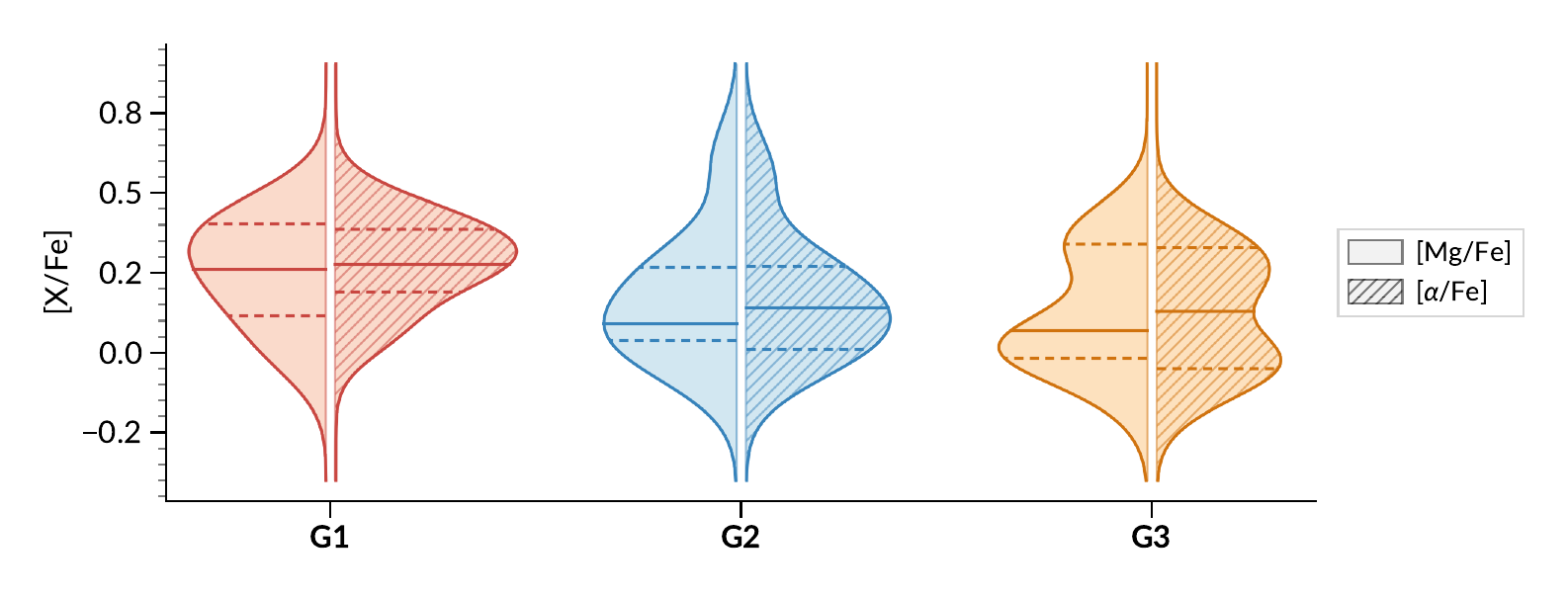}
    \caption{Violin distributions of \mgfe\ (left half) and \afe\ (right half, hatched) for the three GMM groups. Solid horizontal lines mark the median abundance, while dashed lines indicate the 16th and 84th percentiles.}
    \label{fig:alpha_Mg_violin}
\end{figure*}

Figure~\ref{fig:alpha_Mg_violin} shows that the distributions of \mgfe\ and \afe\ are broadly consistent across the three GMM groups, indicating that the use of either quantity does not substantially alter the overall chemical trends discussed in the main text. In G1, \afe\ appears slightly more sharply concentrated around the median than \mgfe, while G2 exhibits very similar distributions for both abundance ratios. In G3, the bimodality already present in \mgfe\ becomes more pronounced in \afe, with the two peaks reaching more comparable relative amplitudes.

\section{Bayesian distances and heliocentric positions of the \tgex\ stars}
\label{append_sec:distances}

\begin{figure*}[ht!]
    \centering
    \includegraphics[width=\linewidth,trim={5mm 7mm 5mm 7mm},clip]{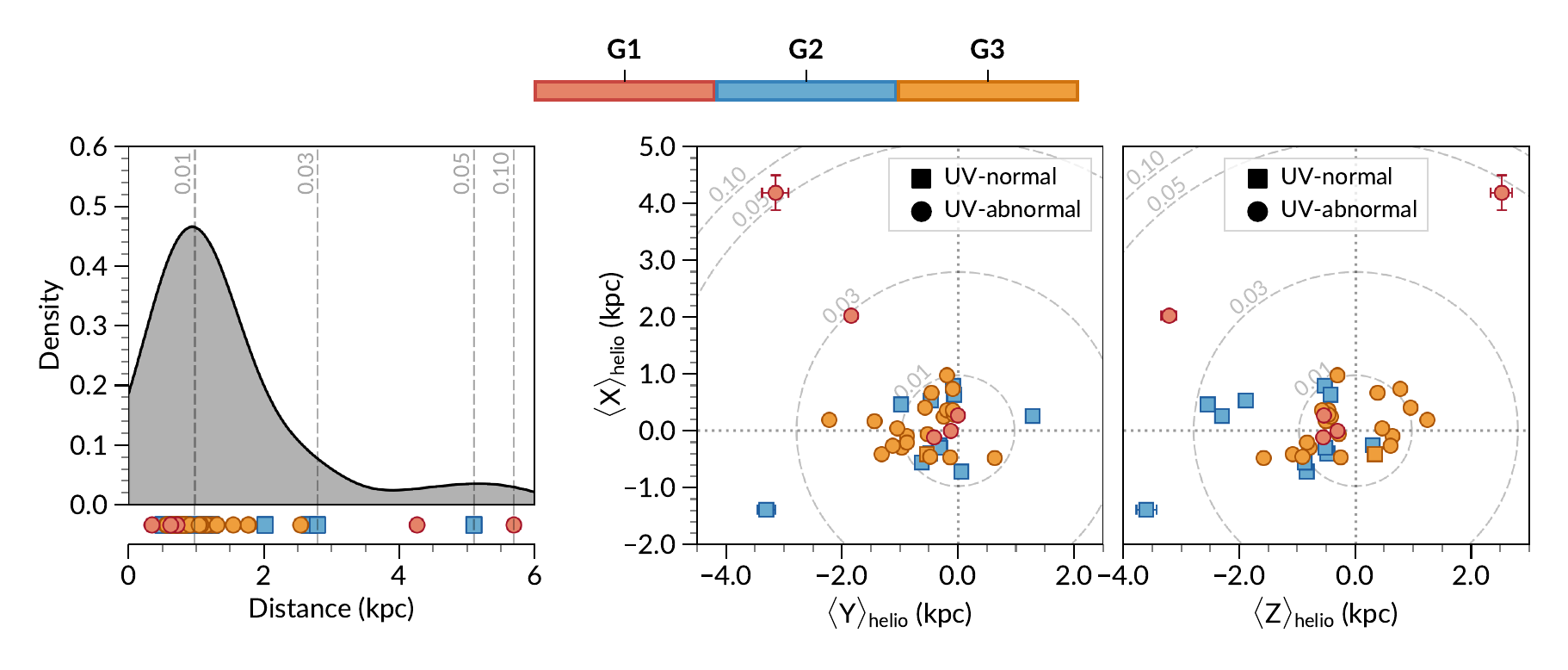}
    \caption{Distance distribution (left) and heliocentric spatial distribution (right) of the 37 \tgex\ stars. The left panel shows the KDE of the Bayesian distances, with the individual stars shown below. The right panels show the heliocentric $\langle X \rangle$--$\langle Y \rangle$ and $\langle X \rangle$--$\langle Z \rangle$ projections, respectively. The dashed concentric circles in the spatial projections and the corresponding vertical dashed lines in the distance panel indicate the maximum distance for which the fractional distance uncertainty satisfies $\sigma_d/d \leq 0.01$, $0.03$, $0.05$, and $0.1$, where $d$ denotes the distance and $\sigma_d$ its corresponding uncertainty. Colours identify the three GMM groups, while squares and circles denote UV-normal and UV-abnormal stars, respectively.}
    \label{fig:distances}
\end{figure*}

Figure~\ref{fig:distances} shows the Bayesian distance distribution ($d$) and heliocentric spatial distribution of the \tgex\ stars. The sample is concentrated relatively close to the Sun, with a median distance of $1.09$~kpc and 34 of the 37 stars lying within $2.8$~kpc. Using the Bayesian distance uncertainties, $\sigma_d$, 16 stars satisfy $\sigma_d/d < 0.01$, 18 additional stars fall in the $0.01 \leq \sigma_d/d < 0.03$ range, and only three objects exceed the $\sigma_d/d \geq 0.03$ threshold. Because the distances were inferred using the Bayesian approach of \citet{Bailer-Jones2015} rather than the direct inversion of the parallax ($\varpi$), the uncertainty behaviour at larger distances is more stable than for simple $1/\varpi$ estimates. The spatial and orbital trends discussed in the main text are therefore dominated by stars with well-constrained distances, while the few more distant objects should be interpreted with additional caution.

\section{Multivariate representation of \tgex\ using `Michinoff' Faces}
\label{append_sec:michinoff}

\begin{figure*}[ht!]
    \centering
    \includegraphics[width=0.9\linewidth,trim={5mm 7mm 5mm 7mm},clip]{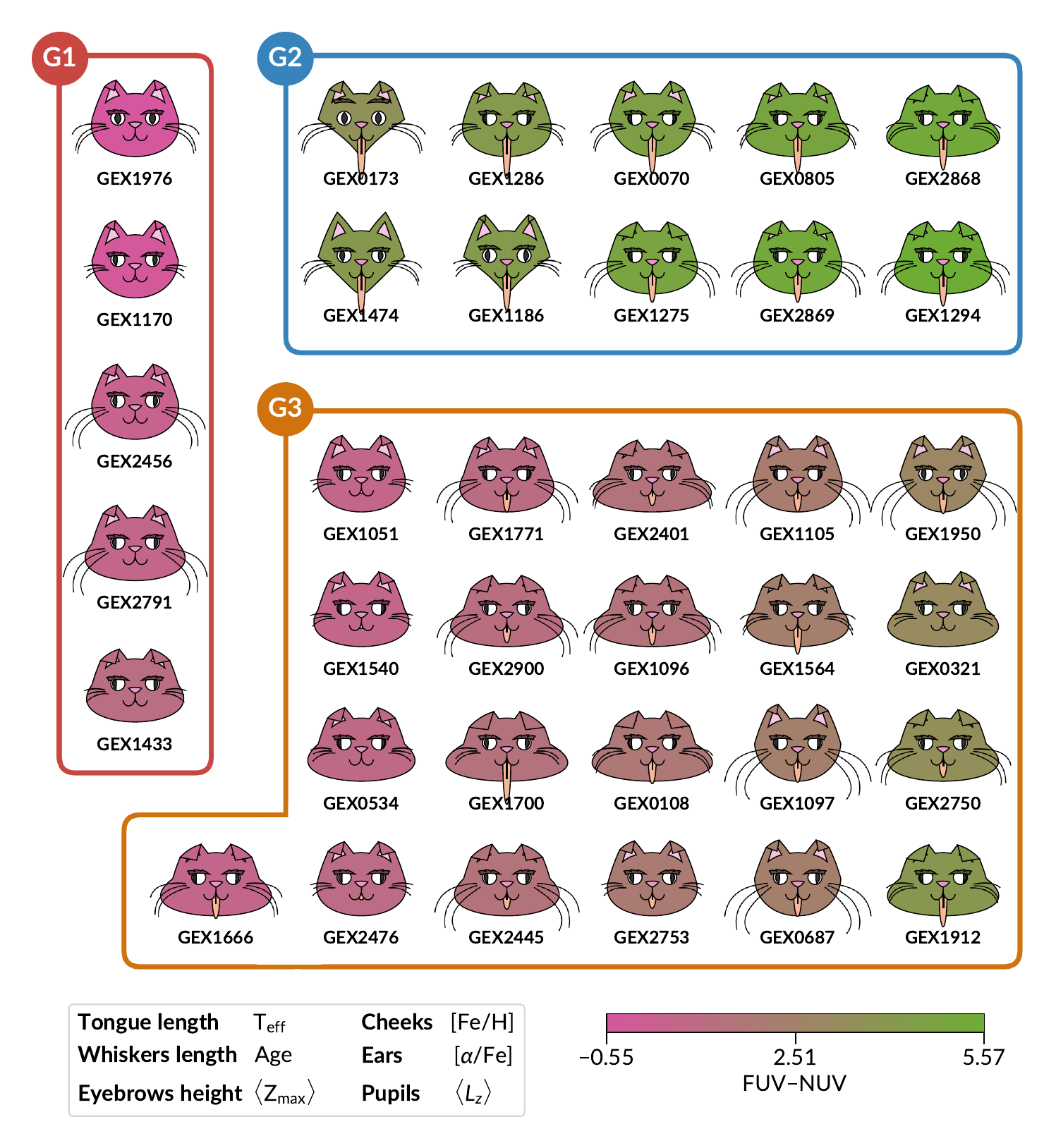}
    \caption{`Michinoff' Faces representation of the \tgex\ sample. 36 out of the 37 \tgex\ stars are shown individually and grouped according to their GMM classification: G1, G2 and G3; one G2 halo star is excluded due to missing $\alpha$-abundance measurements. Each facial feature encodes a different stellar property: eyebrow height represents \zmax, pupil direction represents \lz, tongue length represents \teff, cheek size represents \feh, whisker length represents stellar age, and ear elevation represents \afe. The face colour encodes the \fuvnuv\ colour. Thus, larger values of each parameter correspond to more pronounced features.
}
    \label{fig:all_michinoff}
\end{figure*}

Figure~\ref{fig:all_michinoff} shows the 36 \tgex\ stars represented as `Michinoff' Faces, a multivariate visualization inspired by Chernoff faces \citep{Chernoff1973} and implemented in \cachai\ \citep[see][on the details of this particular module; but see also \citealt{Beltran2025}]{Beltran2026_MichiRNAAS}. The resulting faces visually reproduce the main trends identified for the three GMM groups in \citetalias{Dantas2026_TGEXI} and throughout this paper: G1 is characterised by shorter tongues and more elevated ears, reflecting its lower \teff and higher \afe, while G2 shows longer tongues consistent with a hotter population. G3 displays intermediate and more heterogeneous features, particularly thicker cheeks, reflecting its higher \feh. Notably, GEX0173 stands out from the rest of the sample and corresponds to one of the two halo stars identified in our analysis; the other is not shown due to missing $\alpha$-abundance measurements.

\section{Detailed chemical-abundance patterns of the \tgex\ stars}
\label{append_sec:chemical_patterns}

\begin{figure}[H]
    \centering
    \includegraphics[width=0.85\linewidth,trim={5mm 7mm 5mm 7mm},clip]{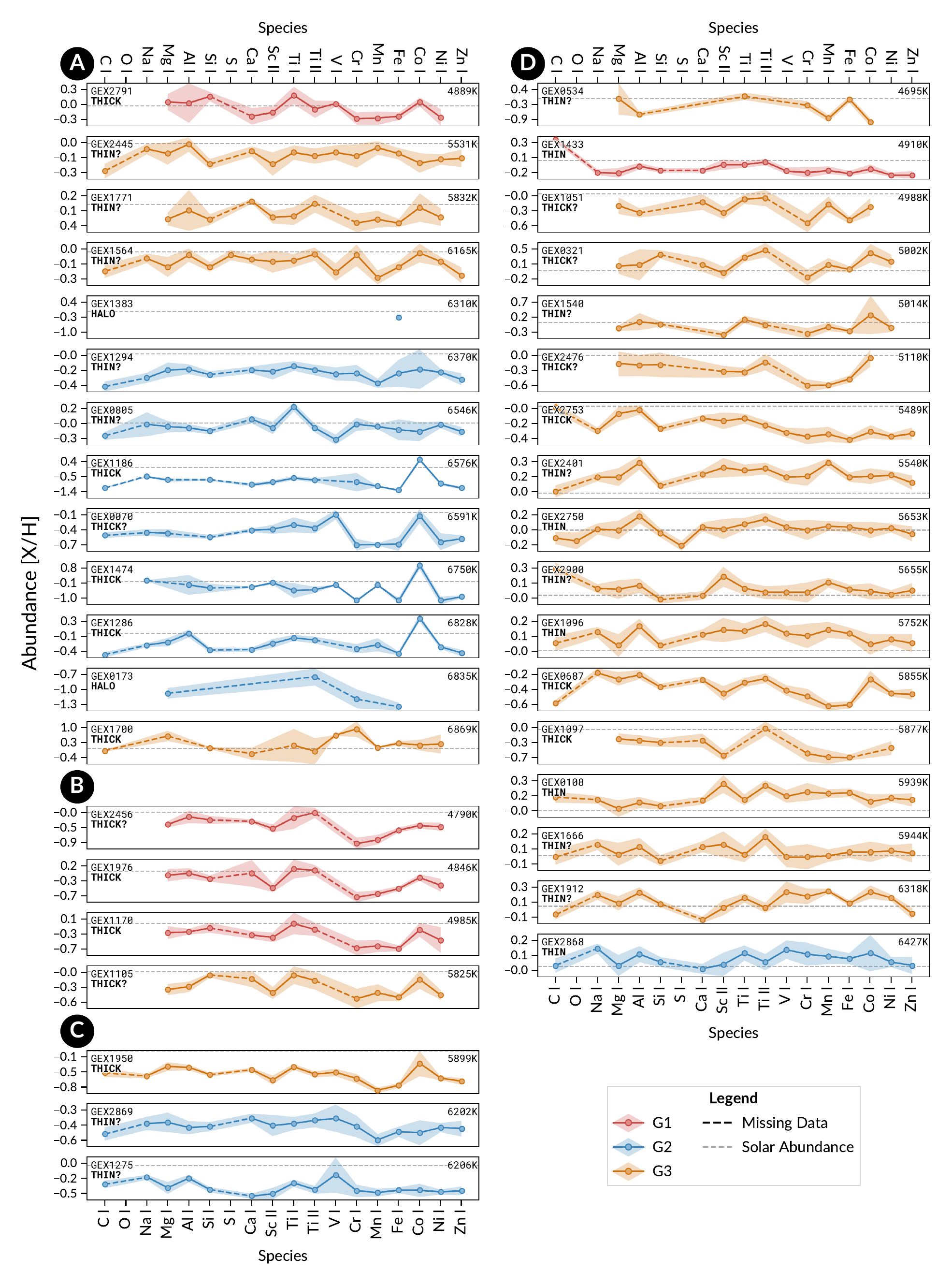}
    \caption{Detailed version of Fig.~\ref{fig:chemical_patterns_simplified}, showing the individual chemical-abundance patterns ([X/H]) for the 37 \tgex\ stars with their $1\sigma$ uncertainties (shaded regions). The HC ordering, \teff\ sorting, GMM colour coding, and line conventions are the same as in the main figure. The horizontal dashed line marks the solar abundance, and each panel includes the stellar identifier, \teff, and Galactic-component classification (\texttt{HALO}, \texttt{THICK}, or \texttt{THIN}). A ``\texttt{?}'' appended to the Galactic-component label indicates an uncertain assignment, corresponding to a membership probability lower than 0.8. The vertical range differs among stars.}
    \label{fig:chemical_patterns}
\end{figure}

\section{Birth- and guiding-radius distributions by thin-disc probability}
\label{append_sec:supplement}

\begin{figure*}[ht!]
    \centering
    \includegraphics[width=\linewidth]{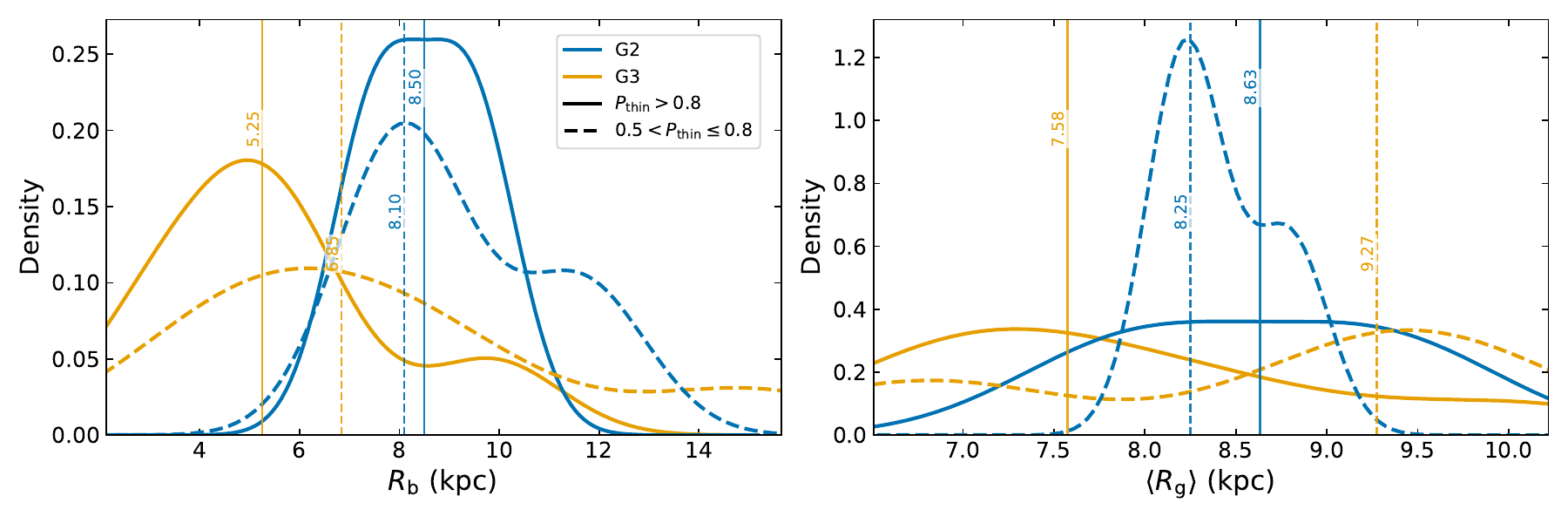}
    \caption{Probability-stratified Gaussian kernel density distributions of the inferred \rb\ (left) and \rgui\ (right) of the probable thin-disc G2 and G3 stars. Colours identify the GMM groups. Solid curves represent stars with $P_{\rm thin}\geq0.8$, while dashed curves represent those with $0.5<P_{\rm thin}<0.8$. Each curve is normalised independently. Vertical lines and numerical labels indicate the corresponding medians. G1 is not shown because only one G1 star satisfies the combined kinematic and chemical selection. This figure stratifies and complements the marginal distributions seen in Fig. \ref{fig:guiding_and_birth_radii} in the main body of the manuscript.}
    \label{fig:rbirth_rgui_kde}
\end{figure*}

Figure~\ref{fig:rbirth_rgui_kde} examines whether the group-level radial differences depend on the adopted chemical thin disc probability. Separate KDEs are shown for stars with $P_{\rm thin}>0.8$ and those with $0.5<P_{\rm thin}<0.8$, after retaining the disc kinematic selection from the Toomre diagram. Each distribution is normalised independently and therefore represents the shape of the corresponding radial distribution rather than the relative number of stars in each subset.

The G3 \rb\ distribution remains shifted towards smaller values than that of G2 in both probability intervals. Among the higher-probability stars, the bootstrap median \rb\ values are $5.25$ and $8.50$~kpc for G3 and G2, respectively. Within G3, the higher-probability subset also has the smaller median \rb\ ($5.25$ versus $6.85$~kpc). For the stars with $0.5<P_{\rm thin}<0.8$, the corresponding medians are $6.85$ and $8.10$~kpc. The persistence of this ordering indicates that the smaller \rb\ of G3 are not produced solely by stars with more marginal chemical thin disc classifications. By contrast, the \rgui\ show no consistent group ordering: the G3 and G2 medians are $7.58$ and $8.63$~kpc in the higher-probability subset, but $9.27$ and $8.25$~kpc in the intermediate-probability subset. The robust distinction between the groups therefore concerns their inferred formation radii rather than their \rgui. Given the small numbers of stars in the individual subsets, the detailed KDE shapes should be regarded as illustrative and the median ordering as the more relevant result. G1 is not shown because only one G1 star satisfies the combined selection and a KDE cannot be estimated from a single object.

\end{document}